\documentclass[a4paper,11pt]{amsproc}
\usepackage{amsmath,amssymb,stmaryrd}
\usepackage{amsfonts}
\usepackage{eucal}
\usepackage{amsthm}
\numberwithin{equation}{section}

\newcommand{\jap}[1]{\langle #1 \rangle}
\newcommand{\beq}{\begin{equation}}
\newcommand{\eeq}{\end{equation}}
\newcommand{\oO}{\omega_0}
\newcommand{\oN}{\omega_N}

\newcommand{\tn}{\mathsf{t}[n]}

\newcommand{\re}{\mathbb{R}}
\newcommand{\ze}{\mathbb{Z}}
\newcommand{\Haar}{\mathsf{Haar}}
\newcommand{\Ha}{\mathsf{H}}
\newcommand{\IP}{\mathsf{P}}
\newcommand{\HD}{\mathsf{D}}
\newcommand{\TV}{\mathsf{V}}
\newcommand{\Fourier}{\mathsf{Fourier}}
\newcommand{\Fo}{\mathsf{F}}
\newcommand{\FH}{\mathsf{FH}}

\theoremstyle{theorem}
\newtheorem{thm}{Theorem}
\newtheorem{cor}[thm]{Corollary}

\theoremstyle{remark}
\newtheorem{rem}{Remark}[section]

\title{Quantization optimized with respect to the Haar basis}
\author{Shu Nakamura}
\address{Department of Mathematics, Faculty of Sciences,
Gakushuin University, 1-5-1, Mejiro, Toshima, Tokyo, Japan 171-8588}
\email{shu.nakamura@gakushuin.ac.jp}

\begin{document}

\maketitle
\vspace{-0.5cm}

\begin{center}
{\small \today}
\end{center}

\begin{abstract}
We propose a method of data quantization of finite discrete-time signals 
which optimizes the error estimate of low frequency Haar coefficients. 
We also discuss the error/noise bounds of this quantization in the Fourier space. 
Our result shows one can quantize any discrete-time analog signal with 
high precision at low frequencies. Our method is deterministic, 
and it employs no statistical arguments, nor any probabilistic assumptions. 
\end{abstract}

\section{Introduction}

It has been known for a long time that the fixed bit depth quantization can encode 
higher bit resolution for low frequencies, possibly by allowing higher level 
of high frequency noise (see, e.g., \cite{LWV}, \cite{OSB} Section~4.9). 
However, as far as the author is aware of, the most research results so far are 
either empirical or based on statistical analysis, which essentially assume 
the signal is a random noise. 
Here we propose a procedure, partly inspired 
by the discrete wavelet theory (see, e.g., \cite{JC}),  to construct 
a quantization of signals, which optimize the error with respect to the Haar coefficients. 
Then we show that this construction also gives good low frequency noise 
bounds in the Fourier space. This argument is completely deterministic, 
and the estimates holds for all the signals without any statistical assumptions. 
We briefly discuss improvements by assuming randomness of the signal 
in Remark~\ref{rem-random}.

We consider analog data on a finite time sequence as follows. 
We set the size of the time sequence as $2^N$, with $N\in\mathbb{N}$. 
We denote the time sequence set by 
\[
\Omega=\biggl\{\tn= -\frac12+\frac{2n-1}{2^{N+1}}\biggm{|} n=1,\dots,2^N\biggr\}\subset[-1/2,1/2]. 
\]
We set $X=\ell^2(\Omega:\re)=\re^\Omega$ as our analog data space, and we denote
\[
f=f[n]=f(\tn)\in [-1/2,1/2], \quad n=1,\dots, 2^N
\]
for $f\in X$. For $f,g\in X$, we introduce an inner product $\jap{f,g}$ by
\[
\jap{f,g}=\jap{f,g}_X= \frac{1}{|\Omega|}\sum_{t\in\Omega} f(t)^*g(t)
=\frac{1}{2^N}\sum_{n=1}^{2^N} f[n]^*g[n]
\]
where $|\Omega|=2^N$. We also denote the \textit{quantized data space}\/ by 
$X_Q=\ell^2(\Omega:\ze)=\ze^\Omega\subset X$. 

\medskip
We now recall the definition of the Haar basis.   
We use the notation:
\[
\hat\Omega = \{(0,1)\}\cup\{(k,j)\mid k=1,\dots, N,j=1,\dots,2^{k-1}\}
\]
is the Haar index sets, and we set the index points by 
\[
\IP[k,j] =  -\frac12+\frac{2j-1}{2^k},\quad (k,j)\in\hat\Omega. 
\]
We set 
\[
\Haar[0,1](t)=1, \quad \Haar[1,1](t)=\begin{cases} 1,\quad &\text{if }0<t<1/2, \\
-1, \quad &\text{if }-1/2<t<0, \\
0,\quad &\text{otherwise}, \end{cases}
\]
and for $k=2,\dots,N$, we set
\beq\label{eq-Haar-def}
\Haar[k,j](t)=2^{(k-1)/2}\Haar[1,1](2^{k-1}(t-\IP[k,j])), \quad 
j=1,\dots, 2^{k-1}.
\eeq
We consider $\Haar[k,j](t)$ as functions on $\Omega$, i.e., $\Haar[\cdot,\cdot]\in\ell^2(\Omega)$. 
It is well-known that the Haar basis
$\{\Haar[k,j]\mid (k,j)\in\hat\Omega\}$
is an orthonormal basis of $\ell^2(\Omega)$.
For $f\in X$, we denote the coefficients of the expansion in the Haar basis by  
\[
\Ha f[k,j] =\jap{\Haar[k,j],f}, \quad (k,j)\in\hat\Omega,
\]
and we write the \textit{Haar transform}\/ of $f$ by 
\[
\Ha f = \{\Ha f[k,j]\}_{k,j}\in \ell^2(\hat\Omega).
\]
We recall $\Ha$ is a unitary map from $X=\ell^2(\Omega)$ to $\hat X=\ell^2(\hat\Omega)$, 
where the inner product on $\hat X$ is defined as usual (without weight):
\[
\jap{\phi,\psi}=\jap{\phi,\psi}_{\hat X}=\sum_{(k,j)\in\hat\Omega} \phi[k,j]^*\psi[k,j],
\quad \phi,\psi\in \hat X.
\]

Now we can state our main result. 

\begin{thm}\label{thm-main}
Let $f\in X=\re^\Omega$. 
Then there exists $g\in X_Q=\ze^\Omega$ such that 
\beq\label{eq-thm-main-2}
|\Ha f[0,1]-\Ha g[0,1]|\leq 2^{-N-1}.
\eeq
and 
\beq\label{eq-thm-main-1}
|\Ha f[k,j]-\Ha g[k,j]|\leq 2^{-N+(k-1)/2},  \quad (j,k)\in\hat\Omega, k\geq 1, 
\eeq
Moreover, 
\beq\label{eq-uniform-error-bound}
\sup_{t\in\Omega} |f(t)-g(t)|\leq 1-2^{-N-1}<1.
\eeq
\end{thm}

\begin{rem}
By the construction in the proof, we learn $g$ is \textit{almost surely}\/  unique, in the sense 
of the Lebesgue measure, and thus the 
choice is optimal in the sense of \eqref{eq-thm-main-2} and \eqref{eq-thm-main-1}. 
If the original data is discrete already, we may add very small noise/randomness 
to make it almost surely unique. This is actually equivalent to employ the randomized choice, 
and the results are independent of the choice of $g$ satisfying these conditions, in any case. 
\end{rem}

\begin{rem}
For $f\in X$, we write the \textit{simple quantization}\/  by 
\[
\bar f[n]= k\in\ze \quad \text{such that } |f[n]-k|\leq \frac12.
\]
Then, naturally, we have 
\[
\sup_{t\in\Omega}|f(t)-\bar f(t)|\leq \frac12,
\]
which is better than \eqref{eq-uniform-error-bound}. On the other hand, we only have 
$|\Ha f[0,1]-\Ha \bar f[0,1]|\leq 1/2$, and 
\[
|\Ha f[k,j]-\Ha \bar f[k,j]|\leq \frac12\cdot 2^{-(k-1)/2},  \quad (j,k)\in\hat\Omega, k\geq 1, 
\]
which are worse than \eqref{eq-thm-main-1} and \eqref{eq-thm-main-2}, especially if $k$ is small. 
\end{rem}

By virtue of \eqref{eq-uniform-error-bound}, we can easily show the following 
result on finite value quantization. 
Let $\ell,m\in\ze$, $\ell+2<m$, and we denote $I=\{\ell,\ell+1,\dots, m\}$, and $\tilde I =[\ell+1,m-1]$. 

\begin{cor}\label{cor-finite-value}
Suppose $f\in X$ with $\mathrm{Ran}[f]=\{f(t)\mid t\in\Omega\}\subset \tilde I$, then 
there is $g\in I^\Omega\subset X_Q$ which satisfies the conditions of Theorem~\ref{thm-main}, 
\eqref{eq-thm-main-2} and \eqref{eq-thm-main-1}. 
\end{cor}

\begin{rem}
Corollary~\ref{cor-finite-value} implies we can allow the analog signal to have values in 
an interval of size $m-\ell-2$. If the resolution $m-\ell$ is smaller or equal to $2$, 
then this result is useless; For example, the so-called one-bit quantization case, 
$m-\ell=1$ and hence $\tilde I=\emptyset$. In such cases, we can still use the method 
combined with a simpler quantization method, for example the pulse width modulation
(PWM) to increase the apparent resolution, and obtain $g$ which enjoy 
the properties \eqref{eq-thm-main-1} for small $k$, but little control for $k$ close to $N$. 
\end{rem}

Now we consider the error/noise of the quantization in the Fourier domain. 
We first fix notation of the discrete Fourier transform. 
We denote the Fourier variable space by 
\[
\Xi = \{-2^{N-1}+1,\dots, 2^{N-1}\} =(-2^{N-1},2^{N-1}]\cap\ze, 
\]
and we write 
\[
\Fourier[\xi](t)=e^{i\oO t\xi}, \quad t\in[-1/2,1/2], \ \xi\in\Xi, 
\]
where $\oO=2\pi$ and $i=\sqrt{-1}$. We denote the discrete Fourier transform by 
\[
\Fo f[\xi] = \jap{\Fourier[\xi],f} =\frac{1}{2^N}\sum_{n=1}^{2^N} e^{-i\oO \tn \xi} f[n]
\]
for $f\in X=\ell^2(\Omega)$. 
The quantization noise in the Fourier space can be estimated as follows: 

\begin{thm}\label{thm-Fourier}
Let $f\in X$ and $g\in X_Q$ as in Theorem~\ref{thm-main}. Then 
\beq\label{eq-thm-Fourier-1}
|\Fo f[0]- \Fo g[0]|\leq 2^{-N-1},
\eeq
and for $\xi\in\Xi$, $\xi\neq 0$,  
\beq\label{eq-thm-Fourier-2}
|\Fo f[\xi]- \Fo g[\xi]|\leq \sum_{k=1}^N 2^{-2N+2(k-1)}
\frac{1-\cos(2^{-k}\cdot2\pi\xi)}{|\sin(2^{-N}\pi\xi)|}. 
\eeq
In particular, 
\beq\label{eq-thm-Fourier-3}
|\Fo f[\xi]- \Fo g[\xi]|\leq \frac{N\pi^2}{2^{N+2}}|\xi|, 
\quad \xi\in\Xi,\xi\neq 0. 
\eeq
\end{thm}

\begin{rem}
Since we have an apriori bound $|\Fo f[\xi]-\Fo g[\xi]|\leq 1$, \eqref{eq-thm-Fourier-3} 
is useful only when $|\xi|\ll 2^N$. If $|\xi|\ll 2^N$, this gives us a very low noise floor, 
i.e., high resolution in the low frequencies.  
If we use the simple quantization, we only have
\[
|\Fo f[\xi]- \Fo \bar f[\xi]|\leq \frac12, \quad \xi\in\Xi, 
\]
and it is not possible to improve it in general. 
\end{rem}

\begin{rem}\label{rem-random}
Our results are completely deterministic, and these inequalities holds for any signal $f\in X$. 
If we suppose the signal is very random, we can use statistical argument to show that 
the quantization noise can be much smaller. 

We recall the argument for the simple quantization case. If $\{f[n]-\bar f[n]\mid n=1,\dots, 2^N\}$ 
are identically distributed independent random variables, then for each $\xi\in\Xi$, we have
\[
\mathbb{E}[(\Fo f[\xi]-\Fo \bar f[\xi])^2] = O(2^{-N}), 
\]
where $\mathbb{E}[\cdot]$ is the expectation, and hence we expect 
$|\Fo f[\xi]-\Fo \bar f[\xi]|=O(2^{-N/2})$. 
This random assumption is unrealistic in applications, and usually high frequency random 
noise (\textit{dithering}) is added to achieve the approximate independence, 
but with higher level of high frequency noise. 

In our setting, if $\{\Ha f[j,k]-\Ha g[k,j]\mid (k,j)\in\hat\Omega\}$ are independent random 
variables, we can show  
\[
\mathbb{E}[(\Fo f[\xi]-\Fo g[\xi])^2] \leq C 2^{-2N}|\xi|^2, 
\]
and hence we expect $|\Fo f[\xi]-\Fo g[\xi]|=O(2^{-N}|\xi|)$, which improves \eqref{eq-thm-Fourier-3}. 
We can also show 
\[
\mathbb{E}[(\Fo f[\xi]-\Fo g[\xi])^2] \leq C 2^{N}|\xi|^{-2}, 
\]
and hence we expect $|\Fo f[\xi]-\Fo g[\xi]|=O(2^{N/2}/|\xi|)$, which is useful 
for $|\xi|\gg 2^{N/2}$. The correlations between Haar coefficients $\{\Ha(f-g)[k,j]\mid (k,j)\in\hat\Omega\}$ 
might usually be smaller than that of $\{f[n]-\bar f[n]\mid n=1,\dots, 2^N\}$, 
but it is not obvious, and we might need 
to add some noise to achieve some approximate independence. 
\end{rem}


\section{Proof of Theorem~\ref{thm-main}}

We first recall the expansion of data in the Haar basis: For $f\in \ell^2(\Omega)$, we have 
\[
f(t) =\sum_{(k,j)\in \hat\Omega} \Ha f[k,j] \Haar[k,j](t), \quad t\in \Omega. 
\]
We write down $\Ha f[k,j]$ more explicitly. We denote the \textit{Haar domains} by 
\[
\HD[k,j] =\biggl(-\frac12+\frac{j-1}{2^k}, -\frac12+\frac{j}{2^k}\biggr)\cap\Omega,
\]
for $k=0,1,\dots,N$,  $j=1,\dots, 2^k$. We note $\HD[N,j]=\{\mathsf{t}[j]\}$, $j=1,\dots, 2^N$. 
We note
\begin{align}
\HD[k-1,j]&=\biggl(-\frac12+\frac{j-1}{2^{k-1}}, -\frac12+\frac{j}{2^{k-1}}\biggr)\cap\Omega
=\biggl(-\frac12+\frac{2j-2}{2^{k}}, -\frac12+\frac{2j}{2^{k}}\biggr)\cap\Omega\nonumber\\
&=\HD[k,2j-1]\cup\HD[k,2j] \label{eq-HD-consistency}
\end{align}
for $k=1,\dots,N$, $j=1,\dots,2^{k-1}$. 
We also note $\Haar[k,j]=\pm 2^{(k-1)/2}$ on $\HD[k,2j]$ and $\HD[k,2j-1]$, respectively, 
and $\Haar[k,j]=0$ otherwise. Thus we have
\beq\label{eq-Haar-formula}
\Ha f[k,j]= 2^{-N+(k-1)/2} \biggl(\sum_{t\in \HD[k,2j]} f(t) 
- \sum_{t\in \HD[k,2j-1]} f(t)\biggr)
\eeq
for $(k,j)\in\hat\Omega$, $k\geq 1$. We also recall 
$\Ha f[0,1] =2^{-N}\sum_{t\in\Omega} f(t)$. 

Inspired by these computations, we define the \textit{total value}\/ of $f\in X$ 
in $\HD[k,j]$ by 
\[
\TV f[k,j]=\sum_{t\in\HD[k,j]} f(t), \quad k=0,1,\dots,N, \ j=1,\dots, 2^k. 
\]
Then we have,  by \eqref{eq-Haar-formula} and \eqref{eq-HD-consistency}, respectively, 
\beq\label{eq-Hg}
\Ha f[k,j]=2^{-N+(k-1)/2}(\TV f[k,2j]-\TV f[k,2j-1]),
\eeq
for $k\geq 1$, and 
\beq
\TV f[k-1,j]=\TV f[k,2j]+\TV f[k,2j-1].\label{eq-TVPg}
\eeq

\medskip
In the following, for a given $f\in X$, we will decide $\TV g[k,j]$ 
inductively on $k$, and show that there exists $g\in X_Q$ 
such that it satisfies these conditions, i.e., we construct $G[k,j]$ 
so that $G[k,j]=\TV g[k,j]$. 
At first, we choose $G[0,1]\in\ze$ so that
\beq\label{eq-H01}
\bigl| \Ha f[0,1] -2^{-N}G[0,1]\bigr|\leq 2^{-N-1}.
\eeq
Such $G[0,1]$ is unique except for a measure zero values of $\Ha f[0,1]$. 

Then, inspired by \eqref{eq-Hg} and \eqref{eq-TVPg}, we find  $G[1,1], G[1,2]\in\ze$ so that 
\beq\label{eq-G1-constraint}
G[1,1]+G[1,2]= G[0,1]
\eeq
and
\[
\bigl| \Ha f[1,1] - 2^{-N} (G[1,2]-G[1,1])\bigr|\leq 2^{-N}. 
\]
Such $G[1,1]$ and $G[1,2]\in\ze$ always exist and almost surely unique (up to measure zero sets), 
since the set of possible $G[1,2]-G[1,1]$ values with 
the condition \eqref{eq-G1-constraint} is $G[0,1]-2\ze$. 
Repeating this procedure, we can construct $G[k,j]\in\ze$,
$k=2,\dots,N$,  $j=1,\dots, 2^{k}$, so that 
\beq\label{eq-Gk-consistency}
G[k-1,j]=G[k,2j]+G[k,2j-1] \quad\text{for }j=1,\dots,2^{k-1}, 
\eeq
and 
\beq\label{eq-HG-error}
\bigl| \Ha f[k,j] - 2^{-N+(k-1)/2} (G[k,2j]-G[k,2j-1])\bigr|\leq 2^{-N+(k-1)/2}
\eeq
for $j=1,\dots, 2^{k-1}$. 
Now we set 
\beq
g[j]=G[N,j], 
\eeq
 for $j=1,\dots,2^{N}$. Then it is easy to show from the above construction, in particular 
by \eqref{eq-TVPg} and \eqref{eq-Gk-consistency}, that 
\[
\TV g[k,j]=G [k,j], \quad k\geq 0, j=1,\dots, 2^k. 
\]
Recalling \eqref{eq-Hg} and \eqref{eq-HG-error}, we learn 
\beq
|\Ha f[k,j]-\Ha g[k,j]|\leq 2^{-N+(k-1)/2}, \quad (k,j)\in\hat\Omega, k\geq 1.
\eeq
The estimate \eqref{eq-thm-main-2} follows immediately from \eqref{eq-H01}.

Now it remains to show the uniform error bound \eqref{eq-uniform-error-bound}. 
Since for each $k$, $\Haar[k,j]$ has disjoint support for $j=1,\dots,2^{N-1}$, 
we have 
\[
\biggl| \sum_{j=1}^{2^{k-1}} (\Ha f[k,j]-\Ha g[k,j])\Haar[k,j](t)\biggr|
\leq 2^{-N+(k-1)/2}\cdot 2^{(k-1)/2}=2^{-N+k-1}
\]
for $k\geq 1$ and $t\in\Omega$. 
By summing up these in $k$, we learn 
\begin{align*}
|f(t)-g(t)| &=\biggl|\sum_{(k,j)\in\hat\Omega} (\Ha f[k,j]-\Ha g[k,j])\Haar[k,j](t)\biggr|\\
&\leq 2^{-N-1}+\sum_{k=1}^N 2^{-N+k-1} =2^{-N-1}+2^{-N}(2^N-1)\\
&=1-2^{-N-1},
\end{align*}
which completes the proof. \qed

\section{Proof of Theorem~\ref{thm-Fourier}}

We first compute the Fourier coefficients of the Haar functions. We denote
\[
\FH[\xi,k,j] = \Fo(\Haar[k,j])[\xi] =\jap{\Fourier[\xi],\Haar[k,j]}
\]
for $\xi\in\Xi$, $(k,j)\in\hat\Omega$. It is obvious that $\FH[0,0,1]=1$; 
\[
\FH[0,k,j]=0\ \text{ for }(j,k)\in\hat\Omega,k\geq 1;  \quad \FH[\xi,0,1] =0\ \text{ for }\xi\neq 0.
\]
In fact, we have $\Ha f[0,1]=\Fo f[0]$. 
In the following, we only consider the case $(k,j)\in\hat\Omega$, $k\geq 1$. 
Recalling the definition of the Haar basis \eqref{eq-Haar-def}, 
we have 
\begin{align*}
\FH[\xi,k,j] 
&= \frac{2^{(k-1)/2}}{2^N}\sum_{n =1}^{2^N} e^{-i\oO \tn \xi}
\Haar[1,1](2^{k-1}(\tn -\IP[k,j]))\\
&=\frac{2^{(k-1)/2}}{2^N}\sum_{n =1}^{2^N} e^{-i\oO (\tn +\IP[k,j])\xi}
\Haar[1,1](2^{k-1}\tn )\\
&= \frac{2^{(k-1)/2}}{2^N}e^{-i\oO \IP[k,j]\xi}
\biggl( \sum_{n =1}^{2^{N-k}} e^{-i\oO\xi 2^{-N}(n -\frac12)}
-\sum_{n =1}^{2^{N-k}} e^{i\oO\xi 2^{-N}(n -\frac12)}\biggr).
\end{align*}
We now compute 
\[
\sum_{n =1}^{2^{N-k}} e^{-i\oO 2^{-N}(n -\frac12)}
=\frac{e^{-i\oN (2^{N-k}+\frac12)\xi}-e^{-i\oN \frac12\xi}}{e^{-i\oN\xi}-1}
=\frac{1-e^{-i\oN 2^{N-k}}}{2i\sin(\oN\xi/2)}, 
\]
where $\oN=2^{-N}\oO$. Similarly, we have 
\[
\sum_{n =1}^{2^{N-k}} e^{i\oO 2^{-N}(n -\frac12)}
=-\frac{1-e^{i\oN 2^{N-k}}}{2i\sin(\oN\xi/2)},
\]
and hence 
\begin{align*}
\sum_{n =1}^{2^{N-k}} e^{-i\oO\xi 2^{-N}(n -\frac12)}
-\sum_{n =1}^{2^{N-k}} e^{i\oO\xi 2^{-N}(n -\frac12)}
&=\frac{2-e^{-i\oN 2^{N-k}\xi}-e^{i\oN 2^{N-k}\xi}}{2i\sin(\oN\xi/2)}\\
=\frac{1-\cos(\oN 2^{N-k}\xi)}{i\sin(\oN\xi/2)}
&=\frac{1-\cos(\oO 2^{-k}\xi)}{i\sin(\oO 2^{-N}\xi/2)}. 
\end{align*}
Thus we obtain 
\[
\FH[\xi,k,j] =\frac{2^{(k-1)/2}}{2^N}e^{-i\oO \IP[k,j]\xi}
\frac{1-\cos(\oO 2^{-k}\xi)}{i\sin(\oO 2^{-N}\xi/2)}, 
\]
and 
\beq\label{eq-FH-exact}
|\FH[\xi,k,j]| = \frac{2^{(k-1)/2}}{2^N}\cdot\frac{1-\cos(\oO 2^{-k}\xi)}{|\sin(\oO 2^{-N}\xi/2)|}
=\frac{2^{(k-1)/2}}{2^N}\cdot\frac{1-\cos(2\pi \cdot 2^{-k}\xi)}{|\sin(2^{-N}\pi\xi)|}, 
\eeq

Now we fix $f\in X$ and its quntization $g$ as in Theorem~\ref{thm-main}, 
and we estimate the bound for $\Fo(f-g)[\xi]$, i.e, the noise bound in 
the Fourier domain. We recall 
\[
\Fo (f-g)[\xi]=\sum_{(j,k)\in\hat\Omega}\FH[\xi,k,j]\Ha(f-g)[k,j], 
\]
and by Theorem~\ref{thm-main}, we know  
\[
|\Ha(f-g)[k,j]|\leq 2^{-N+(k-1)/2}, \quad (k,j)\in\hat\Omega, k\geq 1.
\]
Combining these formulas with \eqref{eq-FH-exact}, we have
\begin{align*}
|\Fo (f-g)[\xi]|
&\leq \sum_{k=1}^N \sum_{j=1}^{2^{k-1}} |\FH[\xi,k,j]|\,|\Ha(f-g)[k,j]|\\
&\leq \sum_{k=1}^N \sum_{j=1}^{2^{k-1}} 2^{-2N+k-1}
\frac{1-\cos(2\pi \cdot 2^{-k}\xi)}{|\sin(2^{-N}\pi\xi)|}\\
&= \sum_{k=1}^N  2^{-2N+2(k-1)}
\frac{1-\cos(2\pi \cdot 2^{-k}\xi)}{|\sin(2^{-N}\pi\xi)|}, 
\end{align*}
which proves \eqref{eq-thm-Fourier-2}. 
We now recall the elementary inequalities: 
\[
|\sin\theta|\geq \frac{2}{\pi}|\theta|
\text{ for }|\theta|\leq\frac{\pi}{2}; 
\quad
1-\cos\theta\leq \frac12\theta^2
\text{ for }\theta\in\re, 
\]
and we have
\[
\frac{1-\cos(2\pi \cdot 2^{-k}\xi)}{|\sin(2^{-N}\pi\xi)|}
\leq \frac{\frac12 (2\pi\cdot2^{-k}\xi)^2}{\frac{2}{\pi}|\pi 2^{-N}\xi|}
=2^{N-2k} \pi^2|\xi|.  
\]
Combining this with  \eqref{eq-thm-Fourier-2},  we have
\[
|\Fo (f-g)[\xi]|
\leq \sum_{k=1}^N 2^{-2N+2(k-1)}\cdot 2^{N-2k} \pi^2|\xi|
=\frac{N\pi^2}{2^{N+2}}|\xi|,
\]
which proves \eqref{eq-thm-Fourier-3}. 
The estimate \eqref{eq-thm-Fourier-1} follows from \eqref{eq-thm-main-1}, 
since $\Fo f[0]=\Ha f[0,1]$ and $\Fo g[0]=\Ha g[0,1]$. 
\qed

\end{document}